# Investigating Central England Temperature Variability: Statistical Analysis of Associations with North Atlantic Oscillation (NAO) and Pacific Decadal Oscillation (PDO)

Jiahe Ling

*Department of Statistics, University of Chicago, Chicago, Illinois*

**ABSTRACT**: This study investigates the variability of the Central England Temperature (CET) series in relation to the North Atlantic Oscillation (NAO) and the Pacific Decadal Oscillation (PDO) using advanced time series modeling techniques. Leveraging the world's longest continuous instrumental temperature dataset (1723–2023), this research applies ARIMA and ARIMAX models to quantify the impact of climatic oscillations on regional temperature variability, while also accounting for long-term warming trends. Spectral and coherence analyses further explore the periodic interactions between CET and the oscillations. Results reveal that NAO exerts a stronger influence on CET variability compared to PDO, with significant coherence observed at cycles of 5 to 7.5 years and 2 to 2.5 years for NAO, while PDO shows no statistically significant coherence. The ARIMAX model effectively captures both the upward warming trend and the influence of climatic oscillations, with robust diagnostics confirming its reliability. This study contributes to understanding the interplay between regional temperature variability and large-scale climatic drivers, providing a framework for future research on climatic oscillations and their role in shaping regional climate dynamics. Limitations and potential future directions, including the integration of additional climatic indices and comparative regional analyses, are also discussed.

**Keywords**: Central England Temperature, Pacific Decadal Oscillation, North Atlantic Oscillation, Time Series Analysis, ARIMA, ARIMAX, Spectral Analysis, Climate Variability

---

## 1. INTRODUCTION

Climate change is widely regarded as one of the most critical challenges of the 21st century, with its impacts becoming increasingly evident through more frequent and intense extreme weather events and profound shifts in global temperature patterns. According to NASA's analysis, 2022 ranks as the fifth warmest year on record since 1880, continuing a trend that has made the past nine years the hottest ever documented (NASA, 2023). This persistent warming trend has been linked to a wide range of global phenomena, including stronger hurricanes, prolonged droughts, and rising sea levels, which collectively pose significant risks to ecosystems, economies, and communities worldwide. Many researchers attribute these changes to anthropogenic greenhouse gas emissions, arguing that human activities such as fossil fuel combustion and deforestation are the primary drivers of climate change (Lashof & Ahuja, 1990). However, some scientists suggest that these trends may be influenced not solely by human activity but also by natural variability within Earth's climate system (Vaidyanathan & Climatewire, 2016). Therefore, gaining a more comprehensive understanding of the mechanisms driving climate variability and change would be highly valuable.

This study aims to explore the role of climate variability in shaping temperature patterns, with a particular focus on understanding how large-scale climatic oscillations influence regional and global temperature trends. Climate variability, driven by both natural and anthropogenic factors, is a key determinant of temperature fluctuations over different timescales. Identifying and disentangling the contributions of natural variability from those of human-induced climate change is critical for accurately predicting future climate scenarios and informing effective adaptation strategies. Among the most influential mechanisms of climate variability are the North Atlantic Oscillation (NAO) and the Pacific Decadal Oscillation (PDO). These large-scale oscillations significantly affect atmospheric and oceanic systems, creating variability that can amplify or offset broader temperature trends. The NAO reflects changes in atmospheric pressure differences between the Icelandic Low and the Azores High, which strongly influence weather and temperature patterns across the North Atlantic and adjacent regions. Positive NAO phases bring stronger westerly winds and milder winters, while negative phases result in colder, drier conditions. Similarly, the PDO is characterized by long-term shifts in sea surface temperature patterns across the North Pacific Ocean, alternating between warm (positive) and cool (negative) phases approximately every 20 to 30 years. These shifts have wide-ranging climatic impacts, influencing temperature, precipitation, and atmospheric circulation far beyond the Pacific region. For example, during positive PDO phases, global temperatures tend to rise more rapidly, whereas negative phases can temporarily moderate global warming, as observed during the "global warming hiatus" in the early 2000s.

The existing body of literature extensively examines the influence of the NAO and PDO on climate variability, emphasizing their critical roles in modulating temperature patterns at both regional and global scales. Building on these studies, this paper aims to reevaluate the effects of the NAO and PDO using advanced analytical techniques, specifically ARIMAX modeling and spectral analysis, applied to the Central England Temperature (CET) dataset. The CET is the longest instrumental temperature record in the world, spanning back to 1659, and provides an unparalleled resource for analyzing long-term temperature variability and trends (Parker et al., 1992). By leveraging these advanced methods and this unique dataset, the study provides a fresh perspective on previously established relationships, offering a refined understanding of how these oscillations drive temperature variability and interact with broader global warming trends. This approach seeks to validate and extend prior findings, contributing to the ongoing discourse on climate variability and its implications for future climate predictions.

This paper is organized as follows. It begins with an in-depth review of existing literature, focusing on the role of the NAO and PDO in influencing climate variability. Next, the data sources utilized and the statistical methodologies, including ARIMAX modeling and spectral analysis, are thoroughly explained. The subsequent section outlines the results of the analysis, highlighting key insights and patterns identified in the context of NAO and PDO impacts. The paper concludes with a discussion of the implications of the findings, an acknowledgment of the study's limitations, and suggestions for future research directions. This work contributes to the field by refining the understanding of NAO and PDO influences on temperature variability, enhancing analytical approaches, and providing valuable knowledge to inform climate policies and mitigation strategies.

## 2. LITERATURE REVIEW

### 2.1　Central England Temperature (CET) Modeling

The CET dataset, the world's longest continuous instrumental temperature record, is pivotal for studying climate variability, long-term trends, and regional dynamics, offering insights into natural variability, anthropogenic impacts, and broader climatic teleconnections.

Studies investigating CET variability and trends have revealed key insights into the dataset's multiscale complexity and long-term properties. Baliunas et al. (1997) applied wavelet analysis to CET data, identifying significant decadal and centennial oscillations, such as 7-8- and 25-30-year cycles (Baliunas et al., 1997). Their findings underscored the intricate interplay between natural climatic forces, such as the cooling associated with the Little Ice Age and subsequent warming phases. Similarly, González-Hidalgo et al. (2020) analyzed CET trends using moving window analysis and Hurst exponent calculations, highlighting the non-linear nature of the dataset (González-Hidalgo et al., 2020). Their findings revealed alternating periods of acceleration and deceleration, challenging the assumption of stationarity and emphasizing the importance of long-term persistence in understanding climate variability.

The CET dataset has also provided a unique perspective on extreme weather events. Brabson and Palutikof (2002) analyzed extreme temperature events using Generalized Pareto Distributions, revealing a decrease in extreme cold winter days linked to atmospheric circulation changes (Brabson & Palutikof, 2002). Conversely, they observed an increase in hot summer days, consistent with global warming. This growing frequency of temperature extremes was further corroborated by Haupt and Fritsch (2022), who employed quantile regression to investigate trends in extreme warm events, emphasizing the heightened variability in recent decades (Haupt & Fritsch, 2022). These studies collectively illustrate CET's utility in documenting changes in extreme weather, which have significant ecological and societal implications.

Explorations of CET's teleconnections with broader climatic phenomena have added depth to its utility. Benner (1999) used spectral and wavelet analyses to uncover a strong decadal relationship between CET and the NAO, while finding no significant connection to the El Niño–Southern Oscillation (ENSO) (Benner, 1999). This study highlighted the influence of regional oscillations on CET variability, underlining the importance of teleconnections in understanding localized climate dynamics.

The role of anthropogenic influences on CET trends has also been a critical area of study. Karoly and Stott (2006) utilized the HadCM3 climate model to attribute a 1°C increase in CET since 1950 to human activities, emphasizing the significant impact of greenhouse gas emissions (Karoly & Stott, 2006). Their findings demonstrated that natural variability alone could not account for the observed warming, underscoring the pressing implications of human-induced climate change. Shi et al. (2022) further identified a pivotal changepoint in the late 1980s, marking a transition to intensified warming consistent with global temperature trends (Shi et al., 2022). These findings solidify CET's position as a reliable indicator of anthropogenic warming on a regional scale.

Seasonal patterns within CET have also been a focus of research, with studies documenting shifts in both amplitude and timing. Proietti and Hillebrand (2017) employed decomposition techniques to examine changes in seasonal cycles, finding reduced amplitude and earlier onset of spring consistent with global warming (Proietti & Hillebrand, 2017). These changes, such as milder winters and advancing seasons, reflect broader climatic shifts that have profound ecological and agricultural implications.

In summary, the CET dataset provides critical insights into climate variability, natural oscillations, anthropogenic impacts, and extreme weather events. Its analyses have highlighted multiscale complexity, teleconnections like the NAO, and shifts in seasonal patterns, solidifying its role as a vital tool for understanding and addressing climate change.

## 2.2 North Atlantic Oscillation (NAO) and Pacific Decadal Oscillation (PDO)

The influence of the NAO and PDO on temperature variability has been widely studied, revealing their significant roles in shaping regional and seasonal temperature patterns.

Castro-Díez et al. (2002) examined the relationship between the NAO and winter temperatures in southern Europe using NCEP/NCAR reanalysis data. Their findings highlighted a non-linear association, where similar NAO index values produced opposing temperature anomalies depending on the location of the NAO centers of action. This underscored the complexity of NAO-driven temperature variability and its regional specificity in southern Europe (Castro-Díez et al., 2002). In Alaska, Papineau (2001) explored the PDO's influence on winter surface air temperatures across 14 stations from 1954 to 2000. The study revealed that warm PDO phases are associated with higher winter temperatures in Alaska, whereas cool phases correlate with colder conditions. This relationship was further modulated by concurrent El Niño or La Niña events, demonstrating the interplay of short-term and long-term climatic drivers on regional temperature patterns (Papineau, 2001). Mantua et al. (1997) analyzed the PDO's impact on climate variability in western North America, identifying a 20–30-year cycle of sea surface temperature (SST) anomalies in the North Pacific. Their findings showed that positive PDO phases aligned with warmer winters in northwestern regions, while negative phases corresponded to cooler temperatures. This study established the PDO as a critical determinant of interdecadal temperature trends in the Pacific basin (Mantua et al., 1997). Benner (1999) utilized spectral and wavelet analyses to investigate NAO-related temperature variability in Central England. The study revealed a strong decadal correlation between CET and the NAO, particularly during winter, demonstrating the NAO's influence on regional climatic variability. However, no significant link was found between CET and ENSO, highlighting the regional specificity of climatic teleconnections (Benner, 1999).

These studies highlight the region-specific impacts of NAO and PDO on temperature variability, as they shape seasonal and interdecadal trends through atmospheric circulation. Their interactions with other drivers, like ENSO, underscore the need for regionally focused research to understand their complexities.

Building on this rich body of research, this study aims to further explore the role of NAO and PDO in shaping CET variability. By employing advanced statistical techniques such as ARIMAX and spectral analysis, this research seeks to reevaluate these oscillations' impacts on CET, offering a deeper understanding of their interactions with regional climate dynamics and contributing valuable insights to the ongoing discourse on climate variability.

## 3. DATA

### 3.1 Overview

The Central England Temperature (CET) dataset, as the world's longest continuous instrumental temperature record, has been extensively studied to understand climate variability and long-term trends. The foundational work by Manley (1974) created the CET dataset, integrating historical temperature records into a consistent series (Manley, 1974). By addressing challenges such as changes in observation sites, urban heat effects, and instrumentation biases, this study set the stage for modern CET analysis. The dataset, spanning from 1659 onwards, revealed long-term fluctuations, including the cooling during the Little Ice Age and subsequent warming trends. Jones and Hulme (1995) extended Manley's dataset, refining it with additional data and corrections for urbanization and station relocations (Hulme et al., 2007). Parker and Horton (2005) addressed uncertainties in CET data for 1878–2003, quantifying random and systematic errors and implementing corrections to maximum and minimum temperatures (D. Parker & Horton, 2005). This study utilizes monthly and annual average CET data from 1723 to 2023, as data prior to 1723 is considered less reliable, and 2024 data remains incomplete (Harvey & Mills, 2003). Seasonal averages are calculated as the mean of relevant monthly data.

The North Atlantic Oscillation (NAO) data, obtained from NOAA, spans 1950 to 2024 (NOAA, 2024a). The NAO index is calculated by projecting the NAO loading pattern onto daily 500-millibar height anomalies over 0–90°N. This loading pattern, derived as the first mode of a Rotated Empirical Orthogonal Function (EOF) analysis using 1950–2000 monthly data, reflects the sea-level pressure difference between the Subtropical High and Subpolar Low. The positive phase indicates lower pressure at high latitudes and higher pressure in the central North Atlantic, influencing the jet stream, storm tracks, and heat transport, leading to temperature and precipitation changes from eastern North America to western and central Europe. The negative phase exhibits the opposite effects.

The Pacific Decadal Oscillation (PDO) data, also sourced from NOAA, spans from 1854 to 2024 (NOAA, 2024b). This study uses monthly and annual average CET data from 1854 to 2023. The PDO index is based on the leading principal component of North Pacific monthly sea surface temperature variability, poleward of 20°N, with global mean temperature removed. Positive PDO phases correspond to warmer waters along the North American coast and cooler waters in the central North Pacific, whereas negative phases exhibit the opposite pattern. These fluctuations significantly influence regional climate, altering temperature, precipitation, and storm patterns across the Pacific Rim and parts of North America.

### 3.2 Trend Visualization

This section presents visualizations of the CET, NAO, and PDO time series to explore their trends and variability. **Figure 1** displays the time series of the CET, NAO, and PDO indices, alongside a combined plot of three series after standardization. The visualizations highlight individual patterns, identify anomalies, and reveal potential connections between temperature and large-scale climatic oscillations, providing a foundation for further analysis.

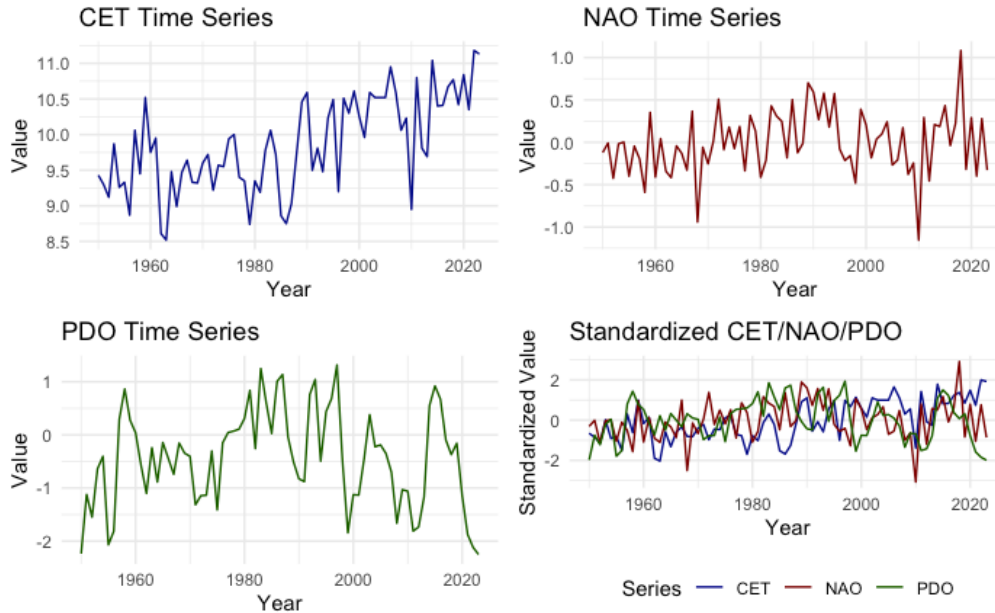

Figure 1 Time Series of CET, NAO, PDO, and Combined Trends.

The original CET time series shows a clear upward trend, particularly pronounced from the late 20th century, reflecting long-term warming consistent with regional and global trends. Short-term fluctuations suggest the influence of natural variability. The NAO time series exhibits strong interannual variability with alternating positive and negative phases but no discernible long-term trend. Prolonged positive phases, such as the mid-1990s, align with milder, wetted winters in Europe, underscoring the NAO's impact on seasonal climate patterns. The PDO time series demonstrates characteristic decadal oscillations, with positive phases dominating from the mid-1970s to mid-1990s and negative phases becoming more frequent in the 21st century, reflecting its role in modulating Pacific Rim climates. The combined standardized plot highlights distinct but occasionally overlapping patterns among CET, NAO, and PDO. While CET shows a clear warming trend, NAO and PDO emphasize interannual and decadal variability. Periods of alignment, such as positive NAO or PDO phases coinciding with CET warming, suggest potential teleconnections that modulate regional climate variability.

The visualizations highlight CET's clear long-term warming trend alongside the interannual variability of NAO and the decadal oscillations of PDO, emphasizing their distinct yet interrelated contributions to regional and global climate dynamics.

### 3.3    Seasonality Visualization

The visualizations in **Figure 2** compare seasonal and annual temperature trends for the CET, NAO, and PDO time series, highlighting distinct patterns for each dataset across seasons (spring, summer, autumn, and winter) and annual averages.

The CET series shows a clear long-term warming trend, especially in summer and winter, with significant variability in winter temperatures influenced by regional factors like atmospheric circulation. Both NAO and PDO time series do not exhibit obvious seasonal difference. In fact, the NAO series exhibits strong interannual variability, particularly in winter, with no distinct long-term trend, reflecting its role in driving seasonal atmospheric dynamics. Besides, the PDO series demonstrates characteristic decadal oscillations with synchronized patterns across all seasons, indicating its broader influence on long-term climate variability with minimal seasonal differentiation.

In summary, the CET series displays distinct seasonal differences, while both the NAO and PDO exhibit more uniform patterns across seasons.

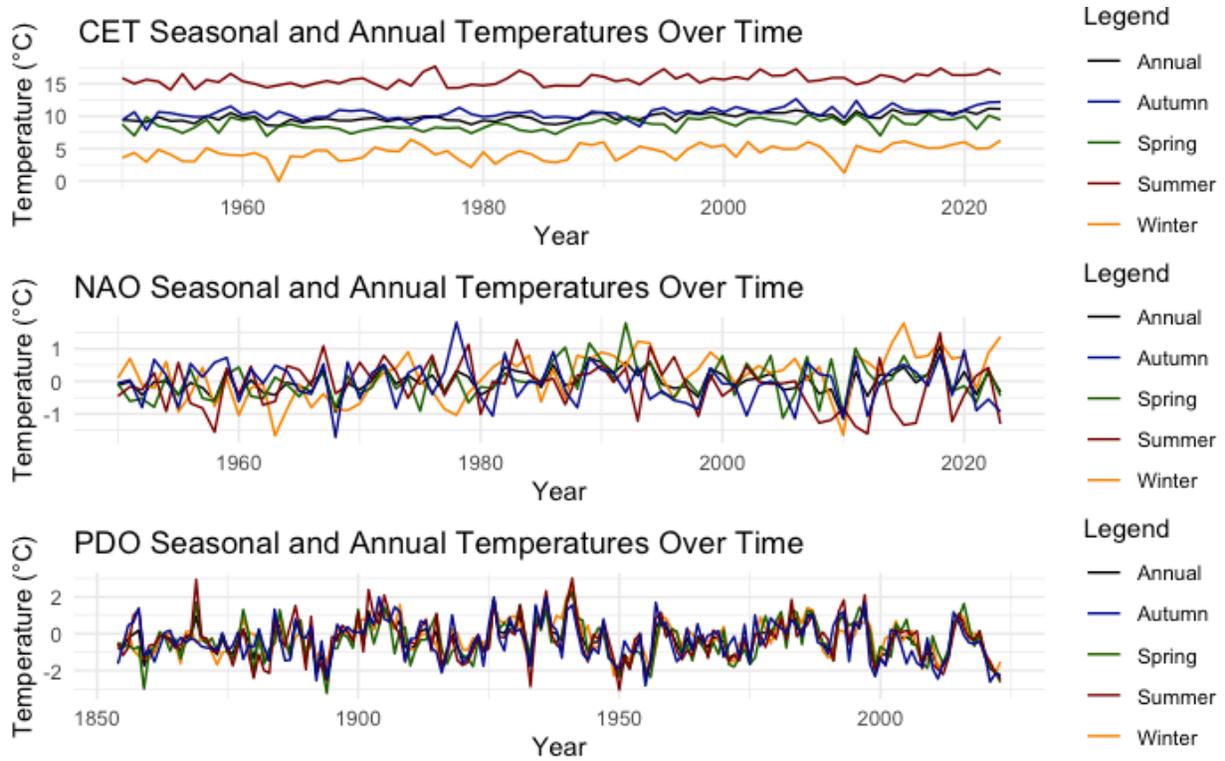

Figure 2 Seasonal and Annual Temperature Trends for CET, NAO, and PDO.

## 4. METHODOLOGY

### 4.1 Modeling the CET Data

To model the CET time series, this study employs an ARIMA(2,1,1) model, a widely used statistical approach for time series analysis that captures both long-term trends and short-term dependencies. The model is designed to capture the dynamics of CET by integrating historical dependencies through autoregressive terms, accounting for random shocks with the moving average component, and ensuring stationarity through differencing.

The ARIMA model is defined by three parameters: $p, d,$ and $q$, representing the order of autoregressive terms, differencing, and moving average terms, respectively. For the CET time series, the general ARIMA(2,1,1) can be expressed as: $y_t - y_{t-1} = \phi_1(y_{t-1} - y_{t-2}) + \phi_2(y_{t-2} - y_{t-3}) + \varepsilon_t + \theta_1 \varepsilon_{t-1}$. The $y_t$ represents the temperature at time $t$, $\phi_1$ and $\phi_2$ are the autoregressive coefficients, $\theta_1$ is the moving average coefficient, and $\varepsilon_t$ is the white noise error term at time $t$ (Shumway & Stoffer, 2017).

The choice of ARIMA(2,1,1) was informed by analyzing the Autocorrelation Function (ACF) and Partial Autocorrelation Function (PACF) of the CET data. The ACF of the original time series shows a slow decay, indicating non-stationarity, while the PACF exhibits a significant spike at lag 1, suggesting the need for differencing.

After applying first differencing, the ACF plot shows a sharp cutoff at lag 1, indicative of a moving average component, and the PACF displays significant spikes at lags 1 and 2, suggesting two autoregressive components. These observations validate the ARIMA(2,1,1) model structure.

The ARIMA(2,1,1) model was fitted to the CET data using Maximum Likelihood Estimation (MLE) to optimize parameter values $(\phi_1, \phi_2, \theta_1)$.

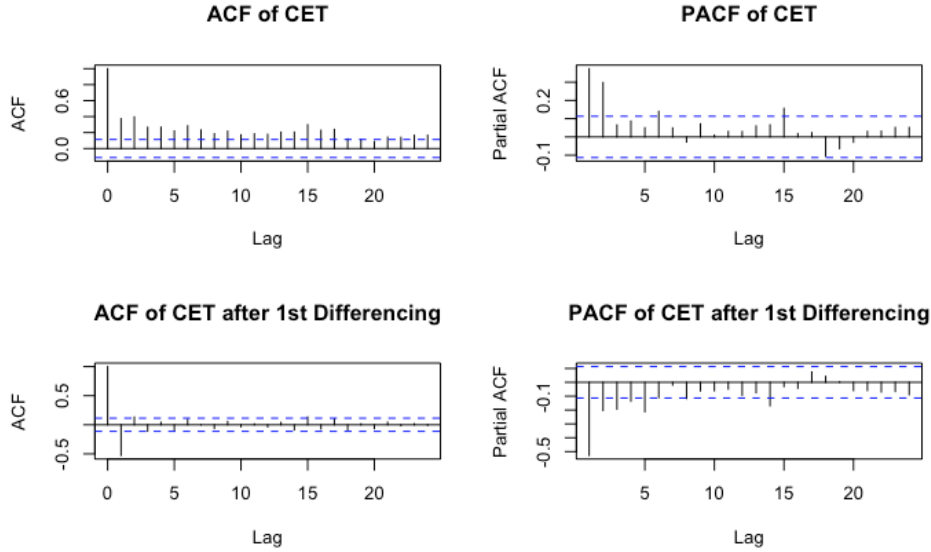

Figure 3 ACF and PACF Plots of CET Before and After First Differencing.

## 4.2 Spectral Analysis

Spectral analysis was performed to investigate the frequency-domain properties of the CET and NAO time series. The goal was to identify dominant periodicities and measure coherence between the two series, providing insights into shared frequency-domain behavior.

Before performing spectral analysis, the stationarity of the CET, NAO, and PDO time series was assessed using the Augmented Dickey-Fuller (ADF) test, which tests the null hypothesis that a time series contains a unit root (non-stationary). For the CET series, the ADF test yielded a p-value of 0.08158, indicating non-stationarity. Similarly, the NAO series had a p-value of 0.244, confirming its non-stationary nature. In contrast, the PDO series showed a p-value of 0.01035, rejecting the null hypothesis and confirming stationarity. To address non-stationarity in the CET and NAO series, first-order differencing was applied, defined as $y'_t = y_t - y_{t-1}$, where $y_t$ is the original series and $y'_t$ is the differenced series. This transformation effectively removed trends and ensured stationarity, as confirmed by subsequent ADF tests.

The analysis utilized the multivariate periodogram, implemented via the spec.pgram function in stats package in R, with a smoothing parameter (LLL) of 9 using the Daniell kernel. The CET and NAO or PDO data were combined into a matrix to compute the spectral density for individual series and the cross-spectral density for paired analysis. The periodogram was used to estimate the squared coherence and associated frequencies, with coherence values indicating the strength of the relationship between CET and NAO or PDO across different frequencies (Shumway & Stoffer, 2017). The coherence is defined as $\rho^2_{y,x}(\omega) = \frac{|f_{xy}(\omega)|^2}{f_{xx}(\omega)f_{yy}(\omega)}$ where $f_{xx}(\omega) = \sum_{h=-\infty}^{\infty} \gamma_{xx}(h)e^{-2\pi i \omega h}$ and $f_{yy}(\omega) = \sum_{h=-\infty}^{\infty} \gamma_{yy}(h)e^{-2\pi i \omega h}$ are the spectral densities of CET and NAO or PDO, respectively, and $f_{xy}(\omega) = \sum_{h=-\infty}^{\infty} \gamma_{xy}(h)e^{-2\pi i \omega h}$ is the cross-spectrum.

To assess the significance of the squared coherence estimates, the critical threshold was computed using the F-distribution. The degrees of freedom $df = 2L\frac{N-1}{N}$ where $L$ is the smoothing width and $N$ is the number of observations. he coherence significance threshold (CCC) was then derived as: $C = \frac{F_{\text{crit}}}{df + F_{\text{crit}}}$ where $F_{\text{crit}} = F(1-\alpha, 2, df-2)$. The significance level $\alpha = 0.01$, and $L = 9$ is commonly used in practice for time series with a moderate number of observations, as it provides a good compromise between stability and detail.

## 4.3   Multivariate Modeling

To explore the relationship between climate oscillations and regional temperature variability, we employed an ARIMAX (Autoregressive Integrated Moving Average with Explanatory Variables) model. The ARIMAX framework is particularly well-suited for time series data as it allows for the inclusion of external regressors. In this case, the NAO and PDO are external regressors to account for potential influences on the CET series.

The ARIMAX(0,1,1) model was applied to annual data spanning from 1951 onward, ensuring consistency in temporal coverage between the CET, NAO, and PDO datasets. The explanatory variables used in the model were the annual indices of NAO and PDO, combined as a matrix of external regressors.

The ARIMAX(0,1,1) model can be expressed as: $y_t = y_{t-1} + \mu + \theta_1 \varepsilon_{t-1} + \beta_1 \text{PDO}_t + \beta_2 \text{NAO}_t + \varepsilon_t$, where $y_t$ represents the temperature at time $t$, $\mu$ is the drift term capturing the overall trend, $\theta_1$ is the moving average coefficient accounting for residual autocorrelation, $\beta_1$ and $\beta_2$ are regression coefficients for PDO and NAO indices respectively, and $\varepsilon_t$ is the white noise error term at time $t$ (Shumway & Stoffer, 2017). The model follows an ARIMA(0,1,1) structure, which includes one differencing step ($d = 1$) to address non-stationarity in the CET series and a single moving average term ($q = 1$) to account for short-term residual autocorrelation. No autoregressive terms ($p = 0$) were included, as no significant lagged relationships for the CET series were observed. This is determined through a grid search process aimed at identifying the model configuration that minimizes the AIC.

Residual diagnostics were conducted to ensure the adequacy of the model fit. These diagnostics included evaluating residuals for randomness and absence of significant autocorrelation.

## 5. RESULTS AND DISCUSSION

### 5.1   CET Modeling

The ARIMA(2,1,1) model was fitted to the annual CET data, providing a robust framework for capturing both short-term fluctuations and long-term trends. The model's coefficients were estimated as follows: $\phi_1 = 0.1108, \phi_2 = 0.1676$, and $\theta_1 = -0.9309$.

The residual variance $\sigma^2 = 0.3396$, with a log-likelihood of -262.03. Model selection metrics, including AIC (532.07), AICc (532.2), and BIC (546.87), confirm the model's suitability for describing the CET time series, balancing complexity with predictive performance. The training set error measures further validate the model's accuracy. A low mean error ($ME = 0.0372$) indicates minimal bias in predictions, while the root mean square error (RMSE = 0.5788) and mean absolute error ($MAE = 0.45$) demonstrate strong performance. Additionally, the mean absolute percentage error ($MAPE = 4.86\%$) suggests that the model's predictions deviate from observed values by less than 5% on average. The minimal autocorrelation in residuals ($ACF1 = 0.00057$) further supports the model's ability to capture the data's temporal structure.

Residual diagnostics confirm the adequacy of the ARIMA(2,1,1) model. The residual plot shows no discernible patterns, indicating independence of residuals. The ACF of residuals falls within the 95% confidence bounds across all lags, suggesting no significant autocorrelation. The histogram of residuals follows a near-normal distribution, further validated by the overlaid density curve. These results demonstrate that the model sufficiently accounts for the variability in the CET data.

The Ljung-Box test provides additional support for the model's validity. With a test statistic ($Q^* = 4.9507$), degrees of freedom of 7, and a p-value of 0.666, the test indicates no significant autocorrelation in the residuals. This suggests that the ARIMA(2,1,1) model has successfully captured the underlying temporal dependencies and stochastic structure of the data.

In summary, the ARIMA(2,1,1) model effectively models the CET annual temperature series, achieving strong predictive accuracy and residual independence. These results highlight the model's reliability for forecasting future temperature trends and its potential for further exploration of climatic variability. The findings underscore the importance of integrating statistical modeling and diagnostic testing to achieve robust insights into time series dynamics.

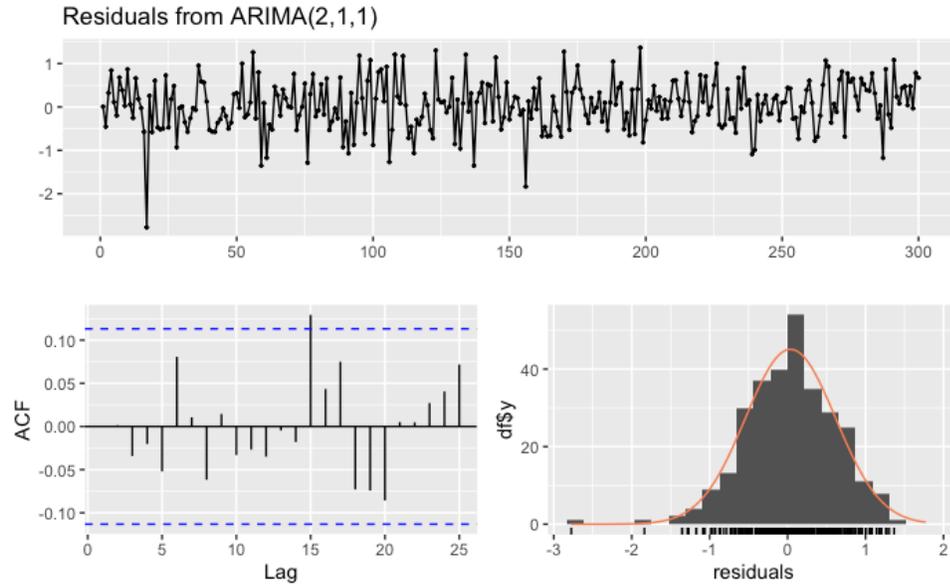

Figure 4 Residual Diagnostics for ARIMA(2,1,1) Model.

## 5.2 Spectral Analysis

The squared coherence analysis between CET and NAO reveals significant frequency bands, indicating strong shared variability between the two time series. Frequencies with coherence values exceeding the critical significance threshold ($C = 0.2606$) were identified as statistically significant. The threshold was determined based on the degrees of freedom (df = 17.75) and a critical F-value ($F_{\text{crit}} = 6.257$) at a 1% significance level ($\alpha = 0.01$). These significant frequencies highlight temporal patterns where CET and NAO exhibit meaningful correlations, supporting the hypothesis of NAO's influence on regional temperature variability.

The significant frequencies span two primary bands: low-frequency (longer period) cycles and high-frequency (shorter period) cycles. The low-frequency band includes frequencies between 0.133 and 0.2, corresponding to periodicities of approximately 5 to 7.5 years. These frequencies align with the NAO's established role as a driver of interannual climatic oscillations, consistent with findings from Benner (1999), who observed strong decadal variability in CET linked to NAO (Benner, 1999). Similarly, the high-frequency band includes frequencies between 0.427 and 0.493, corresponding to periodicities of 2 to 2.5 years. These shorter cycles may reflect localized or seasonal atmospheric dynamics influencing the CET-NAO relationship, potentially linked to sub-annual fluctuations noted in prior studies on atmospheric teleconnections.

The periodogram in **Figure 5** visually highlights these coherence peaks. Notably, the highest coherence occurs near a frequency of approximately 0.15, corresponding to a periodicity of 6.7 years. This finding corroborates the work of Benner (1999), who identified similar interannual patterns in NAO's relationship with regional temperature variability, and González-Hidalgo et al. (2020), who highlighted alternating trends in CET linked to medium-term climatic oscillations (Benner, 1999; González-Hidalgo et al., 2020). The observed coherence patterns underscore the NAO's role in modulating temperature variability at multiple temporal scales, including both interannual and shorter-term dynamics.

The squared coherence analysis between CET and PDO reveals no significant frequencies exceeding the critical significance threshold ($C = 0.2579$). The threshold was determined based on the degrees of freedom (df = 17.94) and a critical F-value ($F_{\text{crit}} = 6.234$) at a 1% significance level ($\alpha = 0.01$). This result indicates an absence of statistically significant coherence between CET and PDO across all examined frequencies.

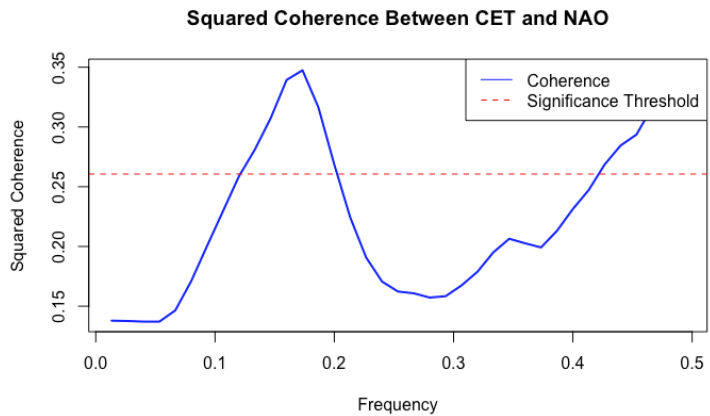

Figure 5 Correlation between Temperature Variability and North Atlantic Oscillation (NAO).

Despite the lack of significant frequencies, the periodogram in **Figure 6** displays prominent peaks in the coherence values, particularly around frequencies corresponding to periodicities of approximately 4 to 5 years and 10 to 12 years. However, these peaks do not exceed the threshold, suggesting that any observed coherence may be due to random noise or weak interactions rather than robust, periodic relationships.

This finding contrasts with the significant coherence observed between CET and NAO, emphasizing the comparatively weaker relationship between CET and PDO. The PDO's influence on regional climates, including Central England, may be less direct than the NAO's, reflecting its primary impact on Pacific Rim and global climate patterns. Previous studies, such as Benner (1999) and González-Hidalgo et al. (2020), have highlighted NAO as a dominant driver of CET variability, while PDO's role remains less evident in this frequency-domain analysis.

The lack of significant coherence between CET and PDO underscores the importance of regional atmospheric and oceanic processes in driving temperature variability. While PDO is a critical factor in modulating global climate patterns, its influence on CET appears to be minimal, suggesting that other factors, such as NAO or localized processes, are more relevant in explaining the variability observed in Central England temperatures. This result provides additional context to the multi-scale climatic drivers impacting CET and highlights the need for further investigation into the interplay between large-scale and regional climate systems.

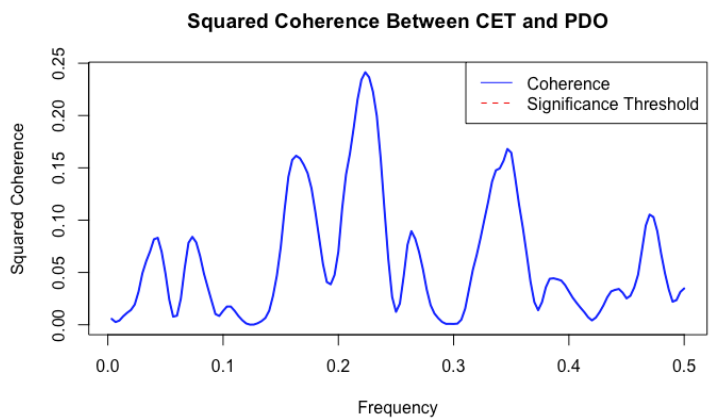

Figure 6 Correlation between Temperature Variability and Pacific Decadal Oscillation (PDO).

## 5.3 Multivariate Modeling

The ARIMAX(0,1,1) model with NAO and PDO indices as external regressors effectively captures the relationship between climatic oscillations and the CET series.

The model's coefficients reveal important insights into the influence of the NAO and the PDO on CET variability. The regression coefficients for PDO ($\beta_1 = 0.0128$) and NAO ($\beta_2 = 0.2835$) indicate that while both oscillations contribute to the temperature variability, the NAO exerts a stronger influence on CET in this context. This aligns with the understanding that NAO directly impacts atmospheric pressure and circulation patterns in the North Atlantic, which are critical drivers of temperature variability in the region. The moving average term ($\theta_1 = -0.8248$) reflects the short-term dependencies in the residuals, effectively capturing the noise structure in the data. The drift term ($\mu = 0.0207$) captures the underlying trend in the differenced series, suggesting a gradual increase in temperature over the observed period, consistent with broader global warming trends.

The residual variance ($\sigma^2 = 0.2444$), along with a log-likelihood of -49.96, indicates the model's robustness in explaining the variability in the CET series. Model selection metrics, including AIC (109.92), AICc (110.83), and BIC (121.3), confirm the suitability of the ARIMAX(0,1,1) model for capturing the dynamics of CET while balancing model complexity and predictive performance.

The training set error measures provide further evidence of the model's accuracy. The low mean error ($ME = 0.0005$) indicates minimal bias in the model's predictions. The root mean square error (RMSE = 0.4772) and mean absolute error ($MAE = 0.3652$) demonstrate the model's strong predictive capability. Additionally, the mean absolute percentage error ($MAPE = 3.77\%$) suggests that, on average, the model's predictions deviate from the observed values by less than 4%, highlighting its effectiveness in capturing the observed variability in the CET series.

Residual diagnostics further validate the adequacy of the ARIMAX(0,1,1) model. The residual plot shows no discernible patterns, confirming the independence of residuals and the absence of structural issues. The ACF of the residuals falls well within the 95% confidence bounds at all lags, indicating no significant autocorrelation. This supports the assumption that the residuals are effectively modeled as white noise. Furthermore, the histogram of residuals closely aligns with a normal distribution, as shown by the overlaid density curve, reinforcing the model's validity under the normality assumption.

The Ljung-Box test provides additional confirmation of the model's adequacy. With a test statistic ($Q^* = 13.881$), degrees of freedom ($df = 9$), and a p-value of 0.1266, the test indicates no significant autocorrelation in the residuals. This suggests that the ARIMAX(0,1,1) model has successfully captured the underlying temporal dependencies and stochastic structure of the data. Finally, the minimal autocorrelation in residuals ($ACF1 = 0.1546$) further supports the model's ability to capture the temporal structure of the CET series.

In summary, The ARIMAX(0,1,1) model highlights that NAO has a stronger influence on CET variability compared to PDO, while also capturing a persistent upward warming trend. The model effectively accounts for both the impact of climatic oscillations and the underlying long-term temperature increase, with robust diagnostics confirming its reliability.

## 6. CONCLUSIONS

This study investigated the variability of Central England Temperature (CET) in relation to two major climatic oscillations, the North Atlantic Oscillation (NAO) and the Pacific Decadal Oscillation (PDO). Leveraging the longest continuous instrumental temperature dataset, this research applied advanced time series modeling techniques, including ARIMA and ARIMAX frameworks, to disentangle the impacts of these oscillations from long-term temperature trends. Spectral and coherence analyses further enhanced the understanding of periodic interactions between CET and the climatic oscillations.

The findings highlight the significant role of NAO in modulating CET variability, with a stronger influence compared to PDO. Additionally, the analysis revealed a persistent upward warming trend in CET, consistent with broader global warming observations. The coherence analysis provided further insights, identifying significant shared variability between CET and NAO across multiple frequency bands, particularly at low frequencies corresponding to cycles of 5 to 7.5 years and high frequencies with cycles of 2 to 2.5 years. These findings underscore NAO's role in driving interannual and shorter-term temperature variability. In contrast, no significant coherence was found between CET and PDO, indicating the absence of robust periodic relationships, despite minor peaks corresponding to cycles of 4 to 5 years and 10 to 12 years, which were not statistically significant.

This study contributes to the literature by integrating multivariate time series modeling and spectral analysis to examine the interplay between regional temperature trends and large-scale climatic drivers. It provides a comprehensive methodology for disentangling the effects of natural variability and anthropogenic influences on long-term temperature records. However, the study is not without limitations. The NAO and PDO indices, while valuable, represent large-scale averages that may not fully capture localized dynamics influencing CET. Furthermore, the exclusion of other potential climatic and non-climatic drivers, such as solar activity or land-use changes, represents a scope limitation.

Future research could explore the integration of additional climatic indices and regional atmospheric data to provide a more nuanced understanding of CET variability. Expanding the analysis to include other regions with long-term temperature records could also offer comparative insights into the influence of climatic oscillations across different climatic zones. Finally, incorporating machine learning techniques into time series modeling may enhance predictive capabilities and reveal hidden patterns in climatic interactions. This study underscores the importance of continued investigation into the intricate relationships between regional temperature variability and global climatic systems in the context of a warming planet.

# 7. REFERENCES

Baliunas, S., Frick, P., Sokoloff, D., & Soon, W. (1997). Time scales and trends in the central England temperature data (1659–1990): A wavelet analysis. *Geophysical Research Letters*, *24*(11), 1351–1354. https://doi.org/10.1029/97GL01184

Benner, T. C. (1999). Central England temperatures: Long-term variability and teleconnections. *International Journal of Climatology*, *19*(4), 391–403. https://doi.org/10.1002/(SICI)1097-0088(19990330)19:4<391::AID-JOC365>3.0.CO;2-Z

Brabson, B. B., & Palutikof, J. P. (2002). The evolution of extreme temperatures in the Central England temperature record. *Geophysical Research Letters*, *29*(24). https://doi.org/10.1029/2002GL015964

González-Hidalgo, J. C., Peña-Angulo, D., & Beguería, S. (2020). Temporal variations of trends in the Central England Temperature series. *Cuadernos de Investigación Geográfica*, *46*(2), 345–369. https://doi.org/10.18172/cig.4377

Harvey, D. I., & Mills, T. C. (2003). Modelling trends in central England temperatures. *Journal of Forecasting*, *22*(1), 35–47. https://doi.org/10.1002/for.857

Haupt, H., & Fritsch, M. (2022). Quantile Trend Regression and Its Application to Central England Temperature. *Mathematics*, *10*(3), 413. https://doi.org/10.3390/math10030413

Hulme, M., Barrow, E., & Atkinson, T. (2007). *Climates of the British Isles: Present, past, and future*. Routledge.

Karoly, D. J., & Stott, P. A. (2006). Anthropogenic warming of central England temperature. *Atmospheric Science Letters*, *7*(4), 81–85. https://doi.org/10.1002/asl.136

Lashof, D. A., & Ahuja, D. R. (1990). Relative contributions of greenhouse gas emissions to global warming. *Nature*, *344*(6266), 529–531. https://doi.org/10.1038/344529a0

Manley, G. (1974). Central England temperatures: Monthly means 1659 to 1973. *Quarterly Journal of the Royal Meteorological Society*, *100*(425), 389–405. https://doi.org/10.1002/qj.49710042511

NASA. (2023, January 12). *NASA Says 2022 Fifth Warmest Year on Record, Warming Trend Continues*. NASA Global Climate Change. https://climate.nasa.gov/news/3246/nasa-says-2022-fifth-warmest-year-on-record-warming-trend-continues/

NOAA. (2024a). *North Atlantic Oscillation (NAO)* [Dataset]. https://www.cpc.ncep.noaa.gov/products/precip/CWlink/pna/norm.nao.monthly.b5001.current.ascii.table

NOAA. (2024b). *The Pacific Decadal Oscillation (PDO)* [Dataset]. https://www.ncei.noaa.gov/pub/data/cmb/ersst/v5/index/ersst.v5.pdo.dat

Parker, D. E., Legg, T. P., & Folland, C. K. (1992). A new daily central England temperature series, 1772–1991. *International Journal of Climatology*, *12*(4), 317–342. https://doi.org/10.1002/joc.3370120402

Parker, D., & Horton, B. (2005). Uncertainties in central England temperature 1878-2003 and some improvements to the maximum and minimum series. *International Journal of Climatology*, *25*(9), 1173–1188. https://doi.org/10.1002/joc.1190

Proietti, T., & Hillebrand, E. (2017). Seasonal Changes in Central England Temperatures. *Journal of the Royal Statistical Society Series A: Statistics in Society*, *180*(3), 769–791. https://doi.org/10.1111/rssa.12229

Shi, X., Beaulieu, C., Killick, R., & Lund, R. (2022). Changepoint Detection: An Analysis of the Central England Temperature Series. *Journal of Climate*, *35*(19), 6329–6342. https://doi.org/10.1175/JCLI-D-21-0489.1

Shumway, R. H., & Stoffer, D. S. (2017). *Time Series Analysis and Its Applications: With R Examples* (4th ed. 2017). Springer. https://doi.org/10.1007/978-3-319-52452-8

Vaidyanathan, G., & Climatewire. (2016, February 25). *Did Global Warming Slow Down in the 2000s, or Not?* SCIAM. https://www.scientificamerican.com/article/did-global-warming-slow-down-in-the-2000s-or-not/